\documentclass[12pt]{iopart}
\usepackage{graphicx}
\usepackage{placeins}
\usepackage{multirow}
\usepackage{tabularx}
\usepackage{booktabs}
\usepackage{url}

\newcommand{\quotes}[1]{``#1''}

\newcommand{\mm}[0]{\,\mathrm{mm}}

\newcolumntype{q}{>{\hsize=.01\hsize}X}
\newcolumntype{s}{>{\hsize=.25\hsize}X}
\newcolumntype{t}{>{\hsize=.33\hsize}X}
\newcolumntype{h}{>{\hsize=.5\hsize}X}
\newcolumntype{z}{>{\hsize=.66\hsize}X}
\newcolumntype{C}{>{\centering\arraybackslash}X}
\newcolumntype{L}{>{\hsize=\hsize \raggedright\arraybackslash}X}
\newcolumntype{R}{>{\hsize=\hsize \raggedleft\arraybackslash}X}
\newcolumntype{H}{>{\hsize=0.5\hsize \raggedright\arraybackslash}X}
\newcolumntype{I}{>{\hsize=0.33\hsize \raggedleft\arraybackslash}X}

\usepackage[dvipsnames]{xcolor}
\definecolor{kitgreen}{RGB}{0, 150, 130}
\definecolor{kitblue}{RGB}{70, 100, 170}
\definecolor{kitblack}{RGB}{0, 0, 0}
\definecolor{kitgray}{RGB}{64, 64, 64}
\definecolor{kityellow}{RGB}{252, 229, 0}
\definecolor{kitorange}{RGB}{223, 155, 27}
\definecolor{kitmaygreen}{RGB}{140, 182, 60}
\definecolor{kitred}{RGB}{162, 34, 35}
\definecolor{kitpurple}{RGB}{163, 16, 124}
\definecolor{kitbrown}{RGB}{167, 130, 46}
\definecolor{kitcyan}{RGB}{35, 161, 224}

\usepackage[pdftex]{hyperref}

\begin{document}

\title[3D-Localization with a Coded Aperture Camera]{3D-Localization of Single Point-Like Gamma Sources with a Coded Aperture Camera}

\author{Tobias~Mei{\ss}ner$^{1, 2^{\ast}}$,
Laura~Antonia~Cerbone$^{3, 4, 5}$,
Paolo~Russo$^{4, 5}$, 
Werner~Nahm$^1$, 
J{\"u}rgen~Hesser$^{2, 6, 7, 8}$}

\address{$^1$ Institute of Biomedical Engineering (IBT), Karlsruhe Institute of Technology (KIT), Karlsruhe, Germany}
\address{$^2$ Mannheim Institute for Intelligent Systems in Medicine (MIISM), Heidelberg University, Mannheim, Germany}
\address{$^3$ Scuola Superiore Meridionale, Napoli, Italy}
\address{$^4$ INFN Sezione di Napoli, Istituto Nazionale di Fisica Nucleare, Napoli, Italy}
\address{$^5$ Dipartimento di Fisica \quotes{Ettore Pancini}, Università di Napoli Federico II, Napoli, Italy}
\address{$^6$ Interdisciplinary Center for Scientific Computing (IWR), Heidelberg University, Heidelberg, Germany}
\address{$^7$ Central Institute for Computer Engineering (ZITI), Heidelberg University, Heidelberg, Germany}
\address{$^8$ CZS Heidelberg Center for Model-Based AI, Heidelberg University, Heidelberg, Germany}

\address{ \hfill \break$^\ast$Corresponding author: \url{publications@ibt.kit.edu}}
\maketitle

\begin{abstract}
\textbf{Objective:} 3D-localization of gamma sources has the potential to improve the outcome of radio-guided surgery. The goal of this paper is to analyze the localization accuracy for point-like sources with a single coded aperture camera.\\
\textbf{Approach:} We both simulated and measured a point-like $^\mathrm{241}$Am source at $17$ positions distributed within the field of view of an experimental gamma camera. The setup includes a $0.11\mm$ thick Tungsten sheet with a MURA mask of rank $31$ and pinholes of $0.08\mm$ in diameter and a detector based on the photon counting readout circuit Timepix3. Two methods, namely an iterative search (ISL) including either a symmetric Gaussian fitting or an exponentially modified Gaussian fitting (EMG) and a center of mass method were compared to estimate the 3D source position. \\
\textbf{Main results:} Considering the decreasing axial resolution with source-to-mask distance, the EMG improved the results by a factor of $4$ compared to the Gaussian fitting based on the simulated data. Overall,  we obtained a mean localization error of $0.77\mm$ on the simulated and $2.64\mm$ on the experimental data in the imaging range of $20$--$100\mm$. \\
\textbf{Significance:} This paper shows that despite the low axial resolution, point-like sources in the nearfield can be localized as well as with more sophisticated imaging devices such as stereo cameras. The influence of the source size and the photon count on the imaging and localization accuracy remains an important issue for further research. The acquired datasets and the localization methods of this research are publicly available on GitHub at \url{https://zenodo.org/records/11449544}.
\end{abstract}

\vspace{2pc}
\noindent{\it Keywords}: gamma imaging, coded aperture, 3D reconstruction, source localization

\section{Introduction}
\label{sec:introduction}
Gamma probes have become an important tool in radio-guided surgery (RGS) for a variety of cancerous diseases over the last few years~\cite{Farnworth2023, Assam2023, Heller2011}.
In contrast to counting probes which provide only an acoustic feedback and a count rate reading, imaging probes, also called gamma cameras allow for a precise detection of structures marked with a radiotracer and additionally give a broader overview of the incision site~\cite{Gonzalez-Montoro2022}.
Mobile gamma cameras are particularly used for sentinel lymph node biopsy (SLNB). An accurate assessment of the axillary lymph node involvement is an essential component in staging breast cancer. Axillary lymph node metastasis is the most important predictor of overall recurrence and survival~\cite{Farnworth2023, Chang2020}. 
Localizing point-like gamma sources in all three spatial dimensions with a mobile gamma camera could provide valuable information to the surgeon~\cite{Bugby2021} and is a first step toward providing valuable depth information in SLNB.

Stereo camera systems are currently under investigation as a potential solution for 3D-localization of gamma sources~\cite{Paradiso2018, Kaissas2015, Kaissas2017, Bugby2021} at the cost of requiring two gamma cameras. An alternative approach is to combine a single mobile gamma camera with external tracking and merge captures from multiple viewpoints into a single 3D map. This approach still requires additional hardware and suffers from increased acquisition time~\cite{Pouw2015}.

A further issue with gamma imaging is the choice of collimation. Most gamma camera systems use either parallel-hole or pinhole collimators to capture an image: while the first allows for high sensitivity, the spatial resolution of resulting images quickly degrades with distance from the source; on the other hand, pinhole collimators offer a poor sensitivity at large distances from the camera. In the 1960s, coded aperture imaging (CAI) was proposed~\cite{Ables1968, dicke1968scatter} as a new imaging technique offering a better trade-off between spatial resolution and sensitivity~\cite{Kulow2020, Accorsi2008, Mu2006}. However, image reconstruction is required to obtain an interpretable image. 

3D imaging of point-like sources is a research field where the capabilities of CAI have not been fully explored yet~\cite{Cannon1979}: by reconstructing the captured detector image at several successive planes (also referred to as in-focus planes), a 3D reconstruction of the source can be computed. The lateral position of the source is encoded by the shift of the mask's shadow, while the source-to-mask distance is related to the size of the shadow. 
In a previous work we were able to show that the most common reconstruction algorithm, called MURA decoding, achieves an axial resolution that is approximately between $15\!:\!1$ and $40\!:\!1$ relative to the lateral resolution~\cite{Meißner2024}. 
This paper aims at answering the following research question: How accurate can a single high-resolution gamma camera with a coded aperture collimator localize a single point-like source in the nearfield setting? 
By localization we mean identifying the 3D coordinates $\left[\hat{x}, \hat{y}, \hat{z} \right]$ in a camera-based coordinate system in the 3D reconstruction 
$\hat{f}\left(x, y, z\right)$ of a single point-like source where the source is assumed to be a isotropically radiating gamma source.
Two different localization methods are investigated: one is based on the center of mass (COM) and the other is an iterative source localization method (ISL) which relies on calculating the deterministic contrast-to-noise ratio (CNR) profile in axial direction. For the depth estimation, two different fitting functions are compared: a Gaussian curve and an exponentially modified Gaussian distribution (EMG). After the performance of the two methods is evaluated on a simulated dataset, the most accurate one is finally used on the experimental dataset we acquired with our gamma camera. In doing so, this paper has the following contributions to the state of the art:

\begin{enumerate}
    \item We show that the CNR profiles of point-like sources are best fitted by an EMG distribution.
    \item We propose an iterative localization method based on a deterministically calculated CNR.
    \item We show that it is possible to achieve a localization accuracy of less than $3\mm$ for point-like sources at a distance of $20$--$100\mm$ obtained, which is comparable to the accuracy of more complex technologies like stereo gamma cameras. 
\end{enumerate}

\noindent To promote transparency and reproducibility, both the acquired datasets and the localization methods of this research are publicly available on GitHub at \url{https://zenodo.org/records/11449544}.

\section{Methods}
\label{sec:mam}
\subsection{Experimental Data}
\label{sec:mam:expdata}
We used a compact gamma camera consisting of a detector and a coded aperture mask to capture the images of a radioactive source. The detector is a hybrid pixel detector (MiniPix EDU, Advacam, Prague, Czech Republic) composed of a $0.5\mm$ thick silicon sensor with a sensitive area of $14.08 \!\times\! 14.08\mm$$^{\mathrm{2}}$, bump bonded to a Timepix3 readout chip  with $256 \!\times\! 256$ pixels of $55 \!\times\! 55\,\,\mu\mathrm{m}^\mathrm{2}$ size, realized by the Medipix3 collaboration at CERN (\url{https://medipix.web.cern.ch/medipix3}). 
The coded aperture mask is composed of a $0.11\mm$ thick Tungsten sheet with a rank $31$ NTHT-MURA pattern with round holes of $0.08\mm$ in diameter~\cite{Accorsi2008}. The basic MURA pattern was duplicated in a $2\!\times\!2$ arrangement leading to as many as $1{,}920$ holes and a total mask size of $9.92\!\times\!9.92\mm$$^{\mathrm{2}}$. 
A 3D-printed case made of acrylonitrile butadiene styrene (ABS) was fabricated to enclose the detector unit and mask, maintaining a detector-to-mask distance $b$ of approximately $20\mm$. The point-like $^\mathrm{241}$Am radioactive source is a sealed sphere of $^\mathrm{241}$Am mainly emitting gamma photons of $59.5$\,keV with a nominal diameter of $1\mm$ with a measured full width at half maximum (FWHM) of $0.65\mm$~\cite{Bertolucci2002}. We used the software \textit{Pixet} provided by the manufacturer of the detector (Advacam: \url{(https://advacam.com/camera/minipix-tpx3}) to capture and save the detector images in the tiff format~\cite{turecek2015pixet}. Pixel values represent the energy deposited in keV integrated over the acquisition time which ranged from $13$ to $15$ minutes.

A total of $17$ images were captured with lateral shifts $y$ from the center at several mask-to-source distances $z$: $0\mm$ for $z=20\mm$; $0\mm$, $2\mm$, $4\mm$, $6\mm$, and $8\mm$ for $z=50\mm$, $z=75\mm$, and at $z=100\mm$ additionally at $y=14\mm$. The $x$-coordinate was kept constant at $0\mm$. Our setup is depicted in Fig.~\ref{fig:img_acq} and a table with all positions can be found in the appendix~\ref{tab:app:17pos}.
We chose these positions for two reasons. First, the two automatic linear axes (Physik Instrumente (PI), Karlsruhe, Germany) available to us imposed practical limitations. Second, it was shown that off-center sources are reconstructed with a lower contrast than centered sources~\cite{Russo2011Evaluation}, however, neither is a difference in the two lateral coordinates reported, nor can we think of a reason to assume an asymmetry. Thus, to spend our measurement time efficiently, we captured only data of a source moving along the positive y-coordinate.

\begin{figure}[tb]
    \centering
    \includegraphics[width=.8\textwidth, trim=0 0 0 0, clip]{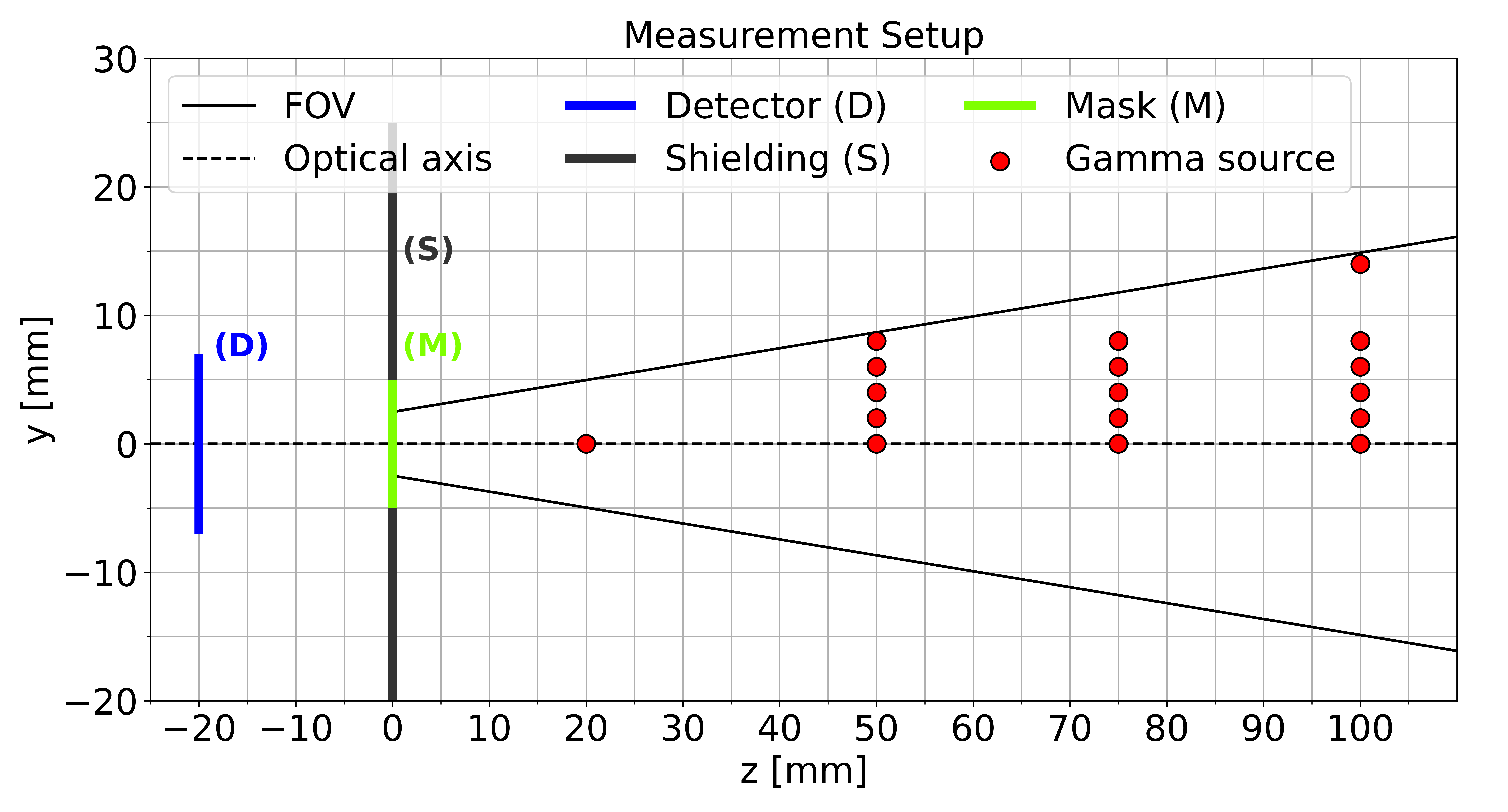}
    \caption{A $^{\mathrm{241}}$Am source (red circle) was placed at $17$ positions within the field of view (FOV) (black lines) of our experimental gamma camera consisting of a detector (D), mask (M) and a shielding (S) at four different mask-to-source distances $z$: $20\mm$, $50\mm$, $75\mm$, and $100\mm$ and varying $y$-coordinates.} 
    \label{fig:img_acq}
\end{figure}

\subsection{Monte Carlo simulation}
\label{sec:mam:mcsim}
We reproduced the experimental setup in silico using the Monte Carlo (MC) simulation software package TOPAS (version 3.8.1)~\cite{Perl2012} which is a wrapper library around Geant4~\cite{Agostinelli2003}. The source was modeled as a homogeneous sphere of $^\mathrm{241}$Am with a $1\mm$ diameter and emitting a total of $10^\mathrm{9}$ photons of $59.5$\,keV. 
We scored all photons that passed the front surface of the detector in a phase space file. From this list, we generated a detector image by computing the 2D histogram of all hits. By doing so, we ignored the charge-sharing effect between neighboring pixels and defective pixels, which is rare in thin silicon-based detectors~\cite{Ruat2012}.  
The resulting gap between simulation and experimental data will be discussed in Sec.~\ref{sec:dis:limits}.

\subsection{Reconstruction}
\label{sec:mam:recon}
Between the two reconstruction methods capable of producing 3D reconstructions, we chose MURA decoding, as it is fast enough to be employed in real-time. For more information about this reconstruction method, the reader is referred to~\cite{Gottesman1989}. Generally, the field of view (FOV) of CAI depends on the reconstruction method used~\cite{Fujii2012}. 
To localize the point-like gamma source, we reconstructed the detector image at many succeeding in-focus planes and thus obtained a 3D reconstruction of the scene. We explored a depth range of $11\mm$ to $130\mm$ with a step of $\Delta z=0.5\mm$, resulting in a stack of $239$ images. Since the size of the reconstructed images depends on $z$, each image of the stack is resized to the maximally occurring image size by bilinear interpolation, which is $254\!\times\!254$ pixels for $z=11\mm$. 
The resizing allows for easy handling of the 3D reconstruction as it can be processed as a 3D matrix, where the $z$-coordinate aligns with the source's extension in $z$-direction. By doing so, the position of the source does not change with respect to the FOV's center, which will be helpful during the localization process.

\subsection{A deterministic contrast-to-noise ratio}
\label{sec:mam:cnr}
As a metric to localize the gamma source, we used the CNR. This was preferred over the reconstructed intensity of the source (which would correspond to the point-spread function (PSF) of the source) since it increases the robustness of the localization procedure and is consistent with the literature~\cite{Meißner2024, Meißner2023, Cerbone2023}. The fitting procedure becomes more robust by using the CNR instead of the PSF, as it calculates a single  CNR value for an entire image stack (the 3D matrix of successive reconstructed planes). This mitigates the contributions from Poisson noise, which is highly correlated along the $z$-axis but remains unconsidered in MURA decoding.  
The CNR is here defined as

\begin{equation}
    {CNR} = \frac{S-B}{\sigma_B}, 
\end{equation}

\noindent where $S$, $B$ and $\sigma_B$ respectively denote the signal as the average of a region of interest (ROI), the background, and the background’s standard deviation which is representative of the image noise.
However, the CNR is usually determined by placing the ROI for the signal and background manually, which is time-consuming and imprecise.
Thus, we compute the CNR automatically under consideration of all possible ROIs. 
We determine $S$ and $B$ by convolution with a kernel, that represents the circular ROI: The kernel consists of a quadratic image twice the size of the ROI's radius in pixels whose values are $0$ everywhere except for the central circular region, where they are set to $1$. The kernel is normalized to the sum of $1.0$, thus the convolution generates an image in which each pixel represents the average intensity of a circular ROI centered at that pixel. 
The convolution is only applied to the central part of the image where the kernel does not extend beyond the boundary of the image. Subsequently, we append rows and columns of NaNs (\quotes{not a number}) to keep the image size constant. 

To determine $\sigma_B$, we apply two convolutions: one to obtain the average intensity of a ROI, just as described above, and another one with the image where each pixel's intensity was squared. We combined both results to obtain the ROI's variance and hence the standard deviation at each possible pixel position.
Finally, we obtain the intensity and the standard deviation at each position for all possible ROIs. To determine the CNR, $S$, $B$, and $\sigma_B$ must be selected: first, we choose $S$ as the highest average intensity and find the position of the corresponding ROIs. All ROIs that overlap with $S$ are then removed. 
Second, the average and the standard deviation of the remaining ROIs are considered as background, and by averaging on them we obtain $B$ and $\sigma_B$ respectively. The result is a CNR value that is independent of manual selection.

\subsection{Localization with the Iterative Source Localization method (ISL)}
\label{sec:mam:localization}

\begin{figure}[tb]
\centering
    \includegraphics[width=1.0\textwidth, trim=0 0 0 0, clip]{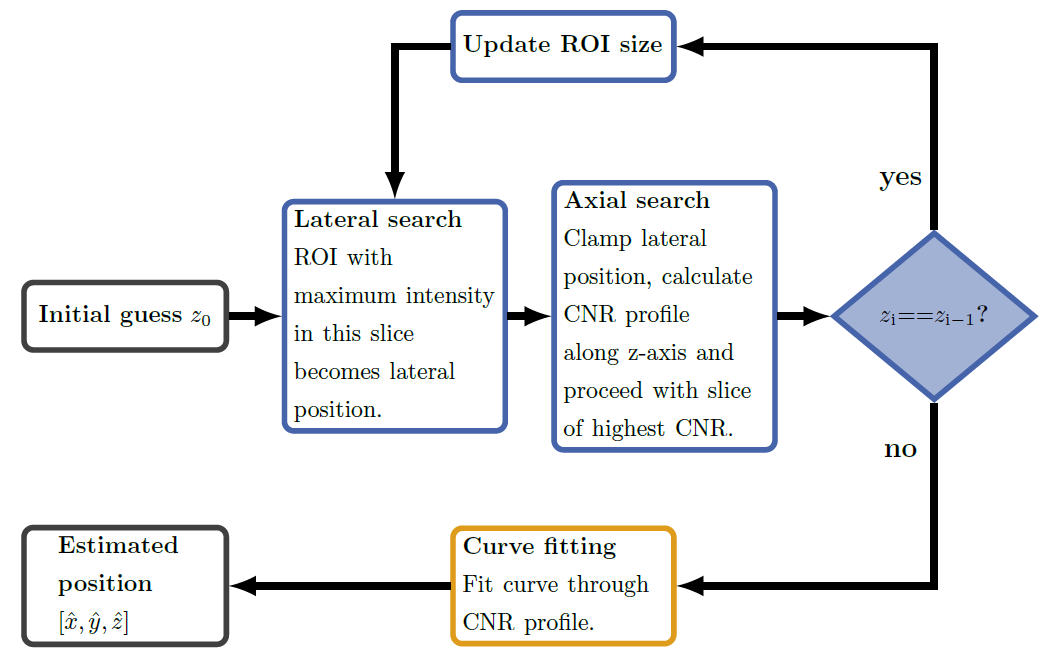}
    \caption{A flowchart representing the Iterative Source Localization algorithm (ISL) with its alternating lateral and axial search and the final curve fitting to estimate the 3D position of the source.}
    \label{fig:isl}
\end{figure}

\begin{figure}[tb]
    \centering
    \includegraphics[width=1.0\textwidth, trim= 6 10 2 4, clip]{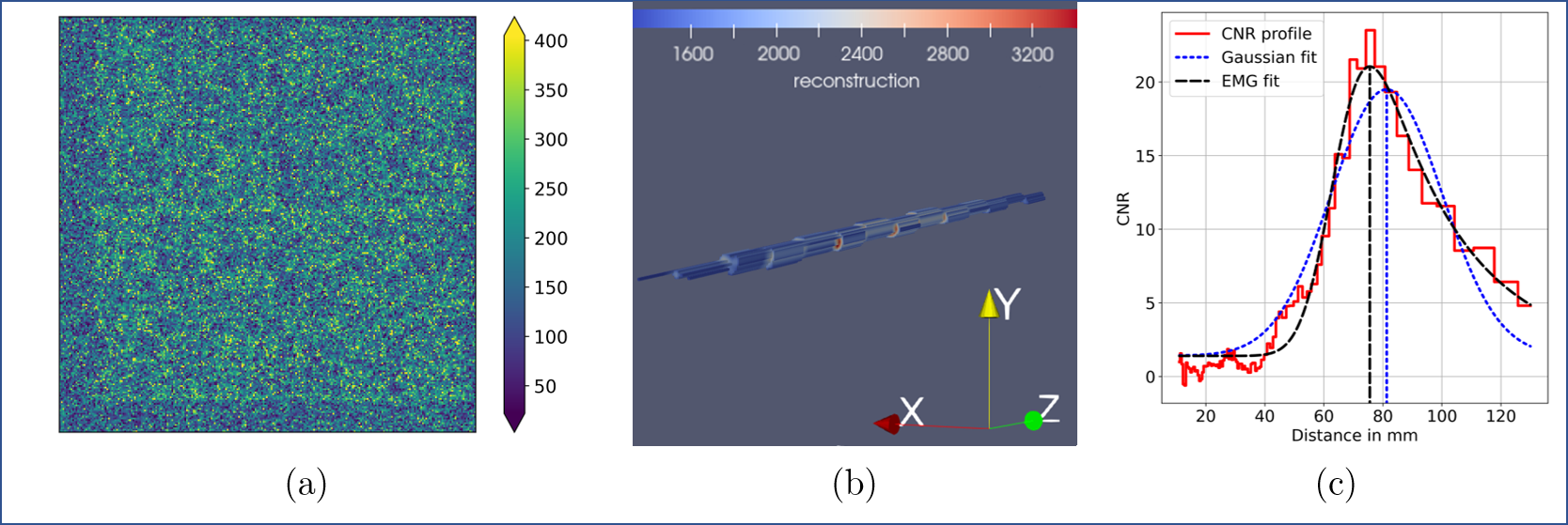}
    \caption{(a) The detector image of the source located at $\left[0, 2, 75\right]\mm$ captured with our experimental gamma camera. The color coding was limited to intensities between the 1$^\mathrm{st}$ and the 99$^\mathrm{th}$ percentile to visualize the projected mask pattern. (b) The 3D reconstruction of the detector image thresholded at the 99$^\mathrm{th}$ percentile to visualize the reconstructed source. Note how the extension in z-direction is much larger than in the lateral directions. (c) The CNR profile with both the Gaussian fit and the EMG fit, resulting in a $z$-estimation of $73.7\mm$ (R$^\mathrm{2}=0.97$) and $81.2\mm$ (R$^\mathrm{2}=0.92$), respectively.} 
    \label{fig:combo_img}
\end{figure}

To obtain the source position from a 3D reconstruction, we adopted the Iterative Source Localization method (ISL):
It consists of iteratively alternating between a lateral and an axial search, followed by fitting a curve into the obtained CNR profile. The algorithm requires as input the reconstructed volume, the FOV for each slice, and an initial guess $z_0$ for the source-to-mask distance. 
The output of the ISL algorithm is the source’s estimated 3D position $\left[\hat{x}, \hat{y}, \hat{z} \right]$ in millimeters. Figure~\ref{fig:isl} shows a flowchart of the ISL algorithm. 

In the first step, i.e. the lateral search, we begin with the image slice that is closest to the initial estimate and begin with the search of the lateral position, i.e. $\left[\hat{x}, \hat{y}\right]$. Similar to the convolution described in Sec.~\ref{sec:mam:cnr}, the average intensity of all possible ROIs is calculated and the position of the highest value is assigned to the current lateral source position. 
The second step is the axial search, where the lateral position is fixed and a CNR profile along the $z$-direction is calculated as described in Sec.~\ref{sec:mam:cnr}. Then, it is checked if a slice with a higher CNR value exists than the current slice at $z_i$. If so, the algorithm starts again with a lateral search followed by an axial search.
The ROI size is updated in each iteration based on the given source size and the FOV of the current slice. 
As soon as $z_i$ does not change within one iteration, the third and last step of the algorithms is entered: The curve fitting. For a more robust and precise depth estimation we apply a curve fitting to the latest obtained CNR profile and analyzed two different fitting functions: once a Gaussian curve with offset $\mathrm{CNR}\left(z\right) = \alpha +\left(\beta - \alpha\right)  \exp\left({-((z-\gamma^2 ))/(2\delta^2 )}\right)$ (in the following denoted as Gaussian) and a scaled exponentially modified Gaussian distribution with offset (EMG)~\cite{Kalambet2011}. 
An EMG distribution emerges from the sum of two independent random variables, where one is normal distributed with mean $\mu$ and variance $\sigma^2$ and the other one exponentially distributed with a rate of $\lambda$. This function was suggested to be potentially a good candidate for the intensity distribution along the axial dimension in Chapter 4.3 in Ref.~\cite{Hellfeld2020}. 
The resulting function to fit the CNR profile has one additional fitting parameter compared to the Gaussian function and can eventually be described as 

\begin{eqnarray}
\mathrm{CNR}\left(z\right) = &\alpha +\left(\beta - \alpha\right) \frac{\lambda}{2}
\mathrm{exp}\left(\frac{\lambda}{2} \left(2\mu + \lambda \sigma^2 - 2z\right)\right) \nonumber \\
& \cdot \mathrm{erfc}\left(\frac{\mu + \lambda \sigma^2 - z}{\sqrt{2}\sigma}
\right),
\end{eqnarray}

\noindent where $\mathrm{erfc}$ represents the complimentary error function that is defined as $\mathrm{erfc}\left(x\right)=2/\sqrt{\pi} \int_x^\infty\!\rme^{-t^2}\,\rmd t$. 
Equations for estimating the peak position (also referred to as \textit{mode}) of an EMG function do exist~\cite{Kalambet2011}, however, for simplicity, we utilize the monomodality, sample the fitted function with a step size of $0.01\mm$ and select the z-value where the sampled function is maximal.
The following initial guesses are directly derived from the CNR profile for the fitting procedure: $\left(\alpha, \beta, \mu, \sigma, \lambda\right) = \left( \mathrm{min}\left(\mathrm{CNR}(z)\right)\!,\, \mathrm{max}\left(\mathrm{CNR}(z)\right)\!,\, \mathrm{argmax}_z\left(\mathrm{CNR}(z)\right)\!,\, 1,\,  1\right)$. The parameter $\lambda$ is dropped for the Gaussian fitting. 
For the remainder of this paper, we will refer to the ISL method with a final EMG fit as ISL-EMG and to the one with a Gaussian fit as ISL-Gaussian. Figure~\ref{fig:combo_img} (c) shows an exemplary CNR profile with both fits. 
After a few iterations and final curve fitting, we eventually obtain the source position $\left[\hat{x}, \hat{y}, \hat{z} \right]$, where the $z$-component is determined by calculating the mode of the fitted function and the $x$ and $y$-component by the brightest ROI position.

\subsection{Localization with the center of mass (COM)}
\label{sec:mam:com}
A more intuitive approach to finding the center of a 3D distribution is with the center of mass (COM). The advantage is that no fitting or user input is required, which makes this method independent from hyperparameters like initial guesses. The COM method relies on the fundamental assumption that the source position is the COM of the largest connected region (LCR) in the given 3D reconstruction. Therefore, we built a pipeline that extracts the LCR and calculates the COM:
first, the 3D reconstruction $\hat{f}\left(x, y, z\right)$ is thresholded by the 99.9$^{\mathrm{th}}$ percentile, a processing step similar to the one proposed by~\cite{Papadimitropoulos2016}.
By thresholding the 3D reconstruction, we assume that all voxels with a lower intensity than the 99.9$^{\mathrm{th}}$ percentile contain background and not the actual reconstructed source. This led to multiple remaining connected regions, i.e. voxel clusters: a large one where we expect the actual source to be located and multiple smaller clusters closer to the camera that are roughly positioned between $11\mm$ and $15\mm$ away from the mask. Hence, as a second step, the region with the largest number of connected voxels is selected and is denoted as LCR: $\hat{f}_\mathrm{LCR}\left(x, y, z\right)$. 
The LCR of the simulated source at at $\left[0, 2, 75\right]\mm$ is depicted in Fig.~\ref{fig:combo_img} (b).
Finally, the COM $\left[\hat{x}, \hat{y}, \hat{z} \right]$ with the intensity of all voxels within this cluster is calculated via the 0$^\mathrm{th}$ ($M_\mathrm{000}$) and 1$^{\mathrm{st}}$ moments ($M_\mathrm{100}$, $M_\mathrm{010}$, $M_\mathrm{001}$) of a 3D distribution, which corresponds to the intensity-weighted mean of $\hat{f}_\mathrm{LCR}$:

\begin{eqnarray}
M_\mathrm{000} &= \displaystyle\sum_{x}\displaystyle\sum_{y}\displaystyle\sum_{z} \hat{f}_\mathrm{LCR}\left(x, y, z\right) \\
M_\mathrm{100} &= \displaystyle\sum_{x}\displaystyle\sum_{y}\displaystyle\sum_{z} x \hat{f}_\mathrm{LCR}\left(x, y, z\right) \\
M_\mathrm{010} &= \displaystyle\sum_{x}\displaystyle\sum_{y}\displaystyle\sum_{z} y \hat{f}_\mathrm{LCR}\left(x, y, z\right) \\
M_\mathrm{001} &= \displaystyle\sum_{x}\displaystyle\sum_{y}\displaystyle\sum_{z} z \hat{f}_\mathrm{LCR}\left(x, y, z\right) \\
\left[\hat{x}, \hat{y}, \hat{z} \right] &= \frac{1}{M_\mathrm{000}}\left[M_{\mathrm{100}} ,M_\mathrm{010} ,M_\mathrm{001}\right] 
\end{eqnarray}

In summary, the COM method performs automatic thresholding, selects the LCR, and determines its COM as the source position. 
Both localization methods are implemented in Python (3.8.18) with NumPy (1.24.4), Tensorflow (2.10.1) for the CNR calculation, SciPy (1.10.1) for the fitting, the VTK library (9.3.0) for finding the LCR and Pandas (2.0.3) for the final analysis. All processing is carried out on a laptop computer with a 6-kernel Intel Core i7-9750H processor (2.6\,GHz), 16\,GB of RAM and a NVIDIA GeForce RTX 2070 with 8\,GB vRAM.

\subsection{Sensitivity analysis}
\label{sec:mam:sa}
While the COM method does not need any input, the ISL method needs an initial guess of the source position by the user. To analyze the impact of the initial guess on the resulting $z$-coordinate, we run the ISL method with several values of the initial $z$-value. To mimic a user scrolling through the slices and selecting the slice where they assume the source is, the initial guess $z_0$ was varied in three categories within a uniform distribution around the true value with $\pm5\mm$, with $\pm10\mm$ and $\pm15\mm$. With the slice interval of $\Delta z=0.5\mm$, the three categories correspond to $\pm10$, $\pm20$, and $\pm30$ slices. Additionally, we analyzed the localization error when we automatically selected the slice with the highest voxel intensity. We applied the ISL method on the simulated dataset with randomly varying initial guesses five times and once with the true $z_0$ and the maximal intensity slice.

\section{Results}
\label{sec:res}
\begin{figure}[tb]
    \centering
    \includegraphics[width=1.0\textwidth, trim=0 0 0 0, clip]{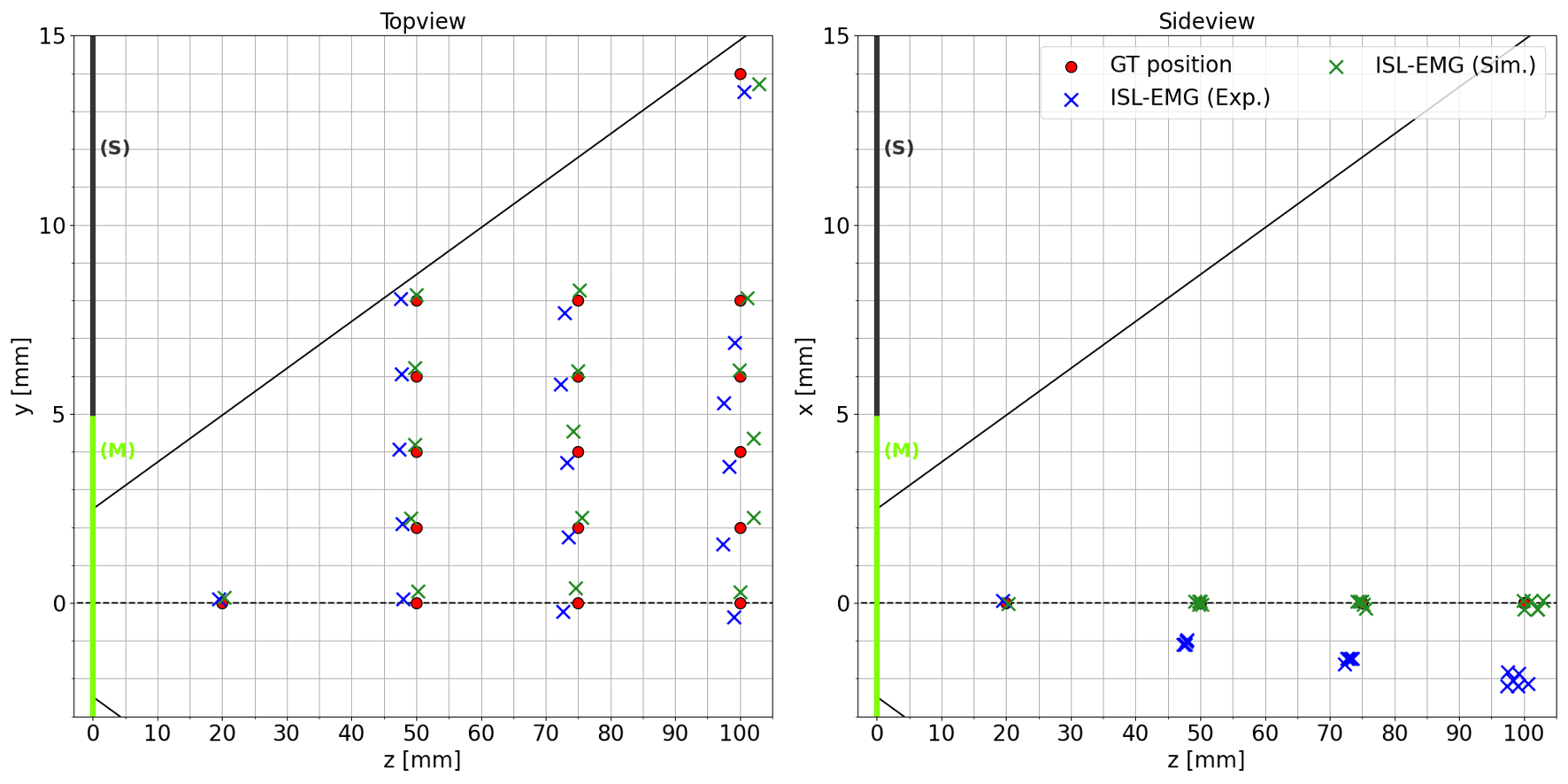}
    \caption{The true source positions (red circles) and the estimates from the ISL-EMG method applied to the simulated (green crosses) and experimental data (blue crosses) in a topview (left) and sideview (right). Note the different range in $z$-direction.} 
    \label{fig:top_side_view}
\end{figure}

\begin{figure}[tb]
    \centering
    \includegraphics[width=1.0\textwidth, trim=0 0 0 0, clip]{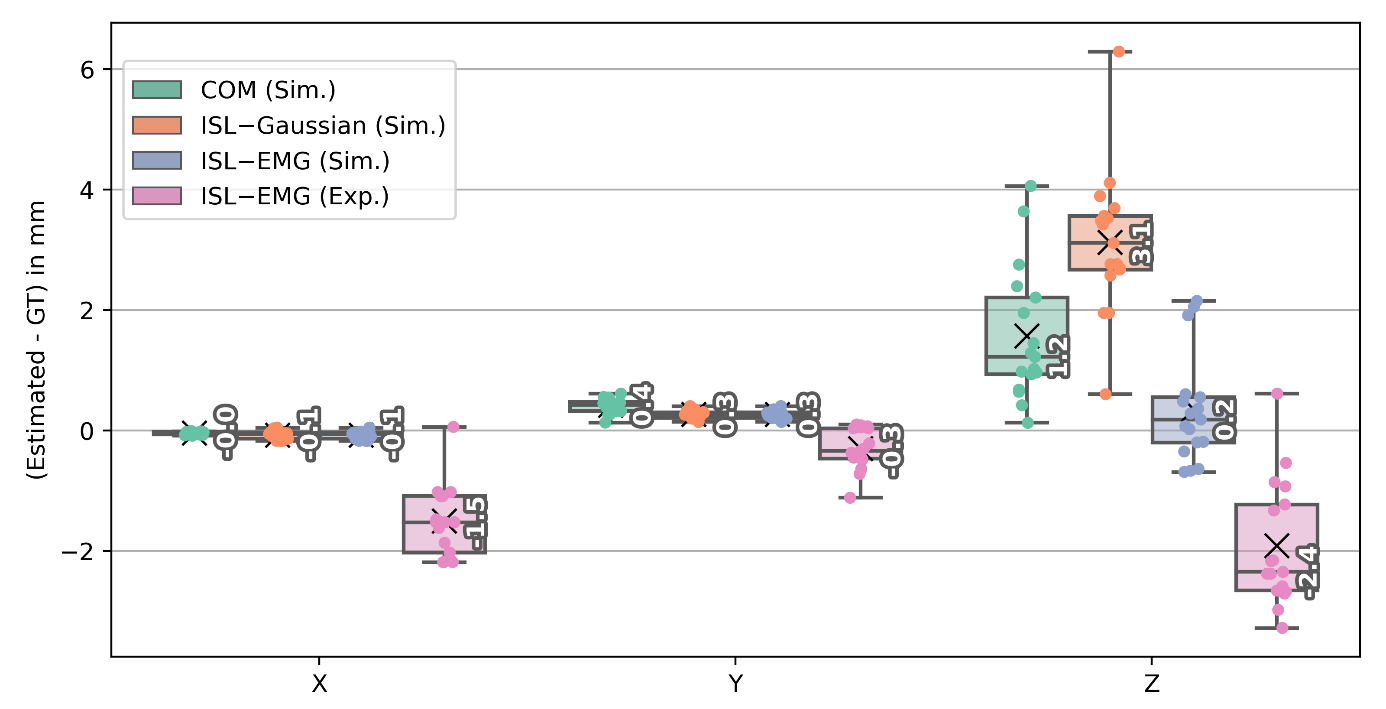}
    \caption{The localization error in mm broken down in $x$, $y$, and $z$-component by localization method (COM, ISL-Gaussian and ISL-EMG) and dataset (simulated and experimental).
    The boxes indicate the 25$^\mathrm{th}$ to 75$^\mathrm{th}$ percentile range, whiskers are maximum and minimum values, and lines are the median error which are also printed vertically in white. The crosses represent the mean values per boxplot.} 
    \label{fig:boxplot}
\end{figure}

\subsection{Simulation results}
\label{sec:res:simres}
The mean localization error and the standard deviation averaged over all $17$ position estimations for both localization methods are presented in Table~\ref{tab:loc_errs}.
The mean localization errors are ($1.65\pm1.05$)\,mm, $(3.13\pm1.15)\mm$ and $(0.77\pm0.62)\mm$ for COM, ISL-Gaussian and ISL-EMG, respectively. 

\begin{table*}[htbp]
\centering
\caption{This table shows the localization errors of the different methods. In addition to the mean error with its standard deviation (STD) over the $17$ 3D reconstructions, also the median error and the relative error of the $z$-component with respect to the true source distance in percent are presented.}
\label{tab:loc_errs}
\begin{tabularx}{1.0\linewidth}{lllrrrrr}
\toprule
\multirow{2}{*}{Data} & \multirow{2}{*}{Method} & \multirow{2}{*}{Fit} & \multirow{2}{*}{Mean} & \multicolumn{2}{r}{Localization error in mm} & \multicolumn{2}{r}{Relative $z$-error in \%} \\
\cmidrule(lr){5-6} \cmidrule(lr){7-8}
 &  &  &  R$^\mathrm{2}$ & Mean$\pm$STD & Median & Mean$\pm$STD & Median \\
\midrule
Sim. & COM & - & - & $1.65\pm1.05$ & $1.34$ & $2.32\pm1.49$ & $1.95$ \\
 & ISL & Gaussian & $0.93$ & $3.13\pm1.15$ & $3.13$ & $4.38\pm1.39$ & $3.90$ \\
 &  & EMG & $0.97$ & $0.77\pm0.62$ & $0.58$ & $0.88\pm0.68$ & $0.60$ \\
Exp. & ISL & EMG & $0.97$ & $2.64\pm0.71$ & $2.59$ & $3.06\pm1.50$ & $2.98$\\
\bottomrule
\end{tabularx}
\end{table*}

A comparison of the localization methods shows that the ISL-EMG method yields overall statistically better results than the COM method.
Using the EMG fit results in a mean coefficient of determination (R$^\mathrm{2}$) of $0.97\pm0.02$ compared to the Gaussian with $0.93\pm0.04$. 
The localization error is larger with ($0.77\pm0.62$)\,mm (EMG) compared to ($3.13\pm1.15$)\,mm (Gaussian) which makes ISL-EMG more accurate than ISL-Gaussian by a factor of $4.06$.
Furthermore, for the ISL-EMG method, the $x$, $y$ and $z$-component contribute on average $3.2$\,\%, $33.6$\,\%, and $63.2$\,\% to the localization error. 
The runtimes on the computer specified above with initialized GPU averaged over the $17$ source positions were ($1.88\pm0.43$)\,s, ($1.48\pm0.25$)\,s, and ($11.76\pm0.46$)\,s for the ISL-Gaussian, ISL-EMG and the COM method, respectively.

\subsection{Sensitivity analysis}
\label{sec:res:sa}
If the initial guess $z_0$ selected for the ISL method was not correct, the mean error remains on average unaffected at ($0.76\pm0.66$)\,mm with random variations of $\pm10$ slices, but increases to ($0.88\pm1.18$)\,mm and ($0.95\pm1.65$)\,mm for larger variations of $\pm20$ and $\pm30$ slices around the true $z$-value. When $z_0$ is chosen according to the voxel with the highest intensity the localization error increases to ($49.85\pm32.28$)\,mm. 
Note that the mean error is calculated based on different numbers of localization errors: While the initial guess with the true slice and with the maximal voxel were calculated once on the dataset, the random guess with increasing variations were carried out five times on the dataset.

\subsection{Experimental results}
\label{sec:mam:expres}
We applied the ISL-EMG method with the ground truth as an initial guess and the correct source FWHM of $0.65\mm$ on the experimental data and a mean localization error of ($2.64\pm0.71$)\,mm was obtained. Especially the errors in $x$ and $z$-direction are not centered around $0$ but at approximately $-1.5\mm$ and $-2.4\mm$ as can be seen in Fig.~\ref{fig:boxplot}. Overall, the localization error is about $3.4$ times worse than obtained from the MC simulation using the same method. 

\section{Discussion}
\label{sec:discussion}
\subsection{Localization method}
\label{sec:dis:localizationmethod}
The MC simulation results reveal that the true mask-to-source distance corresponds to the peak of the CNR profile~\cite{Meißner2023}. As the CNR profile of a single source is skewed towards the camera, the mean value and the mode, i.e. the peak position, do not coincide as they would for symmetric functions. This explains why the COM method overestimates the $z$-coordinate and hence, has a localization error that is approximately $2.1$ times higher than the ISL-EMG method. This emphasizes the importance of considering the positive skewness of the intensity distribution in $z$-direction. 

A major disadvantage of the ISL method we presented, is the required knowledge of the source size and the initial guess $z_0$. One can imagine that the user would quickly scroll through the slices of the 3D reconstructed, spot the source, and draw a circle indicating both the rough source size and the initial guess $z_0$. Nevertheless, it would be more convenient if this process was automated as well. It is imaginable to combine both methods and use the COM to find an estimate for the initialization for the ISL method. 
The sensitivity analysis showed that the mean localization error achieved with variations of up to $30$ reconstructed images is still below $1\mm$, even though the standard deviation increases. 
This can be attributed to the presence of small clusters of high-intensity voxels at close distance to the camera being misidentified as the source.  Additionally, these clusters are the reason why it is not advisable to select the slice with the highest pixel intensity for the initial guess as the large mean error of almost $50\mm$ demonstrates. 
In most cases, these cluster intensities were the highest occurring intensities in the entire 3D reconstruction. This adds another type of systematic artifact caused by MURA decoding besides the known cross-shaped artefact~\cite{Vassilieva2002}, near field effects~\cite{Accorsi2001a, Mu2006}, and ghost source effect~\cite{Willingale1984}: the axial ghost sources. All in all, it can be stated that the sensitivity analysis showed that the ISL method is robust against the initial guess, and the user does not need to find the source position exactly.

The localization errors obtained from the experimental data show a systematic error in the $x$ and $z$-component, despite the overall low error seen in the simulation data. We argue that these systematic errors are due to inaccuracies related to the measurement setup due to the following reasons: 
First, the estimated position in $x$-direction decreases linearly with respect to the mask-to-source distance (see Fig.~\ref{fig:top_side_view}), while it was expected to be zero. This indicates that the source was placed slightly off-center, and additionally, the camera was tilted around the $y$-axis, i.e. pointed upwards. 
Second, the systematic underestimation of the source-to-mask distance is likely caused by an imprecise detector-to-mask distance $b$. 
Even though the camera case was 3D printed, printing tolerances and assembling the case with the Timepix sensor might have led to deviations from the targeted distance of $20\mm$. Furthermore, there is an uncertainty at which depth in the silicon sensor an impinging photon deposits its energy. 
When minimizing the error with respect to the rotation angle $\beta$ and the detector-to-mask distance $b$ from the estimated source positions we obtain an angle and a distance of approximately $1.2$\,degree and $20.6\mm$.

Less surprising is the fact that circa $^{2}\!/_{3}$ of the error stems from the $z$-component, as is in accordance with a previous study, where it was established that with the same camera setup, MURA decoding yields an axial resolution between $15$ and $40$ times worse than the lateral resolution and is degrading with increasing distances~\cite{Meißner2024}. 
The runtimes for the localization methods must be considered in the context of their current implementations which have not been optimized for run time or computational efficiency, but rather represent a proof of concept.

\subsection{Comparison to other localization technologies}
\label{sec:dis:comparison}
To compare our results with other localization technologies we focus on the $z$-component, which dominates the overall accuracy, we set it in relation to the true source-to-mask distance. All distances in this paragraph were converted to source-to-mask distances when necessary. 
Other technologies for localizing gamma sources are either already commercially available~\cite{Pouw2015} or were recently proposed to the research community~\cite{Kaissas2015, Kaissas2017, Paradiso2018, Bugby2021}. The commercial freehandSPECT system was also analyzed in regards to its localization accuracy of small gamma sources and mean errors between $2.9\mm$ and $7.4\mm$ depending on user experience were reported~\cite{Pouw2015}. Though explicit distances are not specified by the authors, from Fig.~2 we can derive a range of $300$-$800\mm$, which results in a relative error below $0.96$\,\% and below $2.46$\,\% for the respective user group. 

Additionally, research groups are investigating stereo gamma cameras with pinhole collimators~\cite{Bugby2021} or coded aperture collimators~\cite{Kaissas2015, Kaissas2017, Paradiso2018, Bugby2021}. Bugby et al. report a median $z$-error of and $1.23\mm$ ($0.83$\,\%) on simulated data and $3.54\mm$ (3.63\%) on experimental data~\cite{Bugby2021}. The research group Kaissas et al. report smaller $z$-errors of $0.28\mm$ ($0.22$\,\%) for a source placed at a source-to-mask distance of $130\mm$ (low rank and thus low-resolution mask) and $1.23\mm$ ($0.94$\,\%) (high-resolution mask)~\cite{Kaissas2015}. The same research group additionally analyzed extended gamma sources of a cylinder with $24\mm$ diameter and $9\mm$ height at increasing source-to-mask distances. Hereby, they show that the localization error deteriorates from $6.1\mm$ to $7.8\mm$ for source-to-mask distances of $140$-$200\mm$~\cite{Kaissas2017}. Paradiso et al.~\cite{Paradiso2018} aim to use their gamma camera at much larger distances with source-to-mask distances between $360\mm$ and $4{,}000\mm$. A $^{\mathrm{241}}$Am source at the distance of $1250\mm$ was estimated to be at $1200\mm$ and at $3000\mm$ to be at $2927\mm$, resulting in localization errors of $50\mm$ ($4.16$\,\%) and $73\mm$ ($2.43$\,\%).

In this paper, with the ISL-EMG method we were able to achieve an error of ($0.88\pm0.68$)\% and ($3.06\pm1.50$)\% on the simulated and on the experimental data with median errors of $0.60$\,\% and $2.98$\,\%. That means for the range of $20$--$100\mm$ source-to-mask distance with a single gamma camera equipped with a coded aperture collimator we obtain a comparable localization accuracy without requiring additional hardware like external tracking or a second camera. 
Stereo cameras which also use coded aperture collimators could benefit from our approach without any changes to the hardware. Their localization procedure, which is currently performed by triangulation, could potentially be made more robust by combining it with two separate single estimations using the here proposed methods.

\subsection{Limitations}
\label{sec:dis:limits}
The study of this paper has a number of limitations that range from specific localization issues to more general difficulties with CAI. First, the algorithm presented can only be applied when no more than one source is located within the FOV.
Second, we only analyzed sources up to a distance of $100\mm$. A degradation of the localization accuracy can be expected, as already shown in ~\cite{Kaissas2017}, which limits the source localization with a single camera to its direct nearfield. In a surgery setting for RGS this could be mitigated by placing the camera closer to the incision site. 
Third, the $0.5\mm$ thick silicon pixel sensor adopted has low detection efficiency at $59.5$\,keV, which implied long acquisition times up to $15$\,min. 
However, Timepix3 detector employing a thick cadmium-zinc-telluride (CdZnTe, or CZT) or cadmium telluride (CdTe) detector exist~\cite{Ruat2012, chmeissani2004first}, which could determine a detection efficiency multiple times higher, correspondingly. 
Based on a rough estimation, the detector images from our MC simulation contain on average $4.8$ times more photons than the actual number of detected photons, which could also partially explain the large difference in localization accuracy. 

This emphasizes that a more precise understanding of the relationship between the number of captured photons and the reconstruction quality and thus the localization accuracy is required, especially in a low-flux real-time application as RGS. Fourth, the localization study with extended gamma sources of~\cite{Kaissas2017} indicate an increasing difficulty to localize sources with increasing size. One possible reason can be found in the supplementary material from~\cite{Fujii2012} (Supplementary Fig.~$7$) where it is shown that the reconstruction quality decreases exponentially with growing source size. Analyzing the influence of the source size was beyond the scope of this paper, but the authors acknowledge that extended sources can represent a serious challenge in the development of a coded aperture camera for RGS. For example, in SLNB lymph nodes cannot be considered as point-like sources as their sizes vary from $5$ to $20\mm$~\cite{Fujii2012, Kaissas2017}, and thus, the influence of the source size on the source localization is yet to be investigated. 

In this study, we used a single camera setup to analyze the 3D-localization accuracy of a point-like source. However, we expect other factors to affect the accuracy, including the detector-to-mask distance which directly influences the magnification of the mask pattern. We further assume the pinhole size, pinhole shape (round or square) and the MURA rank to affect the accuracy as well as the size of the source. These factors were beyond the scope of this work, but their investigation could lead to a broader understanding of 3D-localization.

Due to practical restrictions in our experimental setup we did not evaluate the localization accuracy beyond $100\mm$. However, the advantage of our approach is that the camera can be positioned in close proximity to the source (up to around $11\mm$ for the given detector-to-mask distance), which allows for the capture of sources that may fall outside the FOV of stereoscopic cameras.
Theoretically other reconstruction methods exist, e.g. 3D-MLEM~\cite{Mu2006}, however the runtime reported is in the range of multiple minutes which renders it unsuitable for the real-time usage in RGS.

\section{Conclusion \& Outlook}
\label{sec:candc}
In this paper, we investigated the question of how accurately a single gamma camera with a coded aperture collimator is able to localize in 3D single point-like sources in its nearfield. Our proposed algorithm iteratively searches for the source position based on a rough initial estimate. Mean localization errors below $3\mm$ based on experimental data and even below $1\mm$ based on data obtained by MC simulation were achieved. We showed that when localizing point-like sources, it is important to consider the decreasing axial resolution which manifests in a positively-skewed CNR profile. Incorporating this behavior in form of an EMG fitting increases localization accuracy by an average factor of $4$ compared to using a standard Gaussian fitting. The occurrence of systematic localization errors in the experimental data emphasizes the importance of a thorough assembly and calibration process in the measurement setup. 

For future research the following issues need to be addressed in order to leverage the full potential of CAI in RGS: The most used radiotracer in nuclear medicine is $^{\mathrm{99m}}$Tc which emits gamma photons of higher energy than was investigated in this paper~\cite{Peterson2011}.
A larger and thicker mask ($1\mm$ thick with $0.25\mm$ pinholes) for $^{\mathrm{99m}}$Tc imaging has already been designed and is currently being tested by the authors; further phantom studies will also be conducted in the near future. Furthermore, it is conceivable to train a machine learning algorithm to estimate the source position and potentially together with the source size. 
A better understanding and a solution to the problem of imaging extended sources poses an important milestone on the way to a fast and high-resolution gamma camera.

\newpage
\section*{Supplementary information}
\subsection*{List of abbreviations:}
ABS: acrylonitrile butadiene styrene,
CAI: coded aperture imaging, 
CNR: contrast-to-noise ratio, 
FWHM: full width at half maximum, 
FOV: field of view, 
MURA: modified uniformly redundant array, 
NTHT: no-two-holes-touching, 
PSF: point-spread function, 
ROI: region of interest, 
SLNB: sentinel lymph node biopsy, 

\subsection*{Declarations}
\subsubsection*{Ethics approval and consent to participate}
Not applicable.

\subsubsection*{Competing interests}
The authors declare that they have no competing interests.

\subsubsection*{Consent for publication}
Not applicable.

\subsubsection*{Availability of data and material}
The datasets generated, the localization methods and code for the analysis are publicly available on Github at \url{https://zenodo.org/records/11449544}.

\subsubsection*{Funding}
This project was partly funded by Zentrales Innovationsprogramm Mittelstand (ZIM) under grant KK5044701BS0 from the German Federal Ministry for Economic Affairs and Climate Action. Open access funding was provided by the library of the Karlsruhe Institute of Technology. An additional part of the funding was provided by the INFN Naples within the Medipix4 project.  

\subsubsection*{Authors' contributions}
Contributions are listed according to the CRediT system. TM: conceptualization, formal analysis, investigation, methodology, software, validation, visualization, and writing (original draft). LAC: conceptualization, and writing (original draft). PR: conceptualization, project administration, resources, supervision, and writing (review and editing). WN: conceptualization, funding acquisition, supervision, and writing (review and editing). JH: conceptualization, funding acquisition, project administration, resources, supervision, and writing (review and editing). All authors read and approved the final manuscript.

\subsubsection*{Acknowledgements}
The authors would like to acknowledge the MEDIPIX collaboration (\url{https://medipix.web.cern.ch/home}) and the INFN which, as a member of Medipix2 and Medipix4 Collaborations at CERN, granted use of the MinipixEDU Timepix3 detector used in this work. We, additionally, acknowledge the data storage service SDS@hd supported by the Ministry of Science, Research and the Arts Baden-W{\"u}rttemberg (MWK) and the German Research Foundation (DFG) through grant INST 35/1314-1 FUGG and INST 35/1503-1 FUGG and support by the KIT-Publication Fund of the Karlsruhe Institute of Technology. We acknowledge the use of DeepL Write BETA (\url{https://www.deepl.com/de}) to improve the language of the final draft.

\FloatBarrier
\appendix
\setcounter{section}{1}

\begin{table*}[htbp]
\centering
\caption{We both simulated and captured detector images of a $^\mathrm{241}$Am radioactive source with a nominal diameter of $1\mm$ emitting mainly gamma photons of $59.5$\,keV at $17$ positions given in millimeters within the nearfield of our gamma camera.}
\label{tab:app:17pos}
\begin{tabularx}{1.0\linewidth}{@{}rrrrrrrrrrrrrrrrrr@{}}
\toprule
\textbf{} & \textbf{1} & \textbf{2} & \textbf{3} & \textbf{4} & \textbf{5} & \textbf{6} & \textbf{7} & \textbf{8} & \textbf{9} & \textbf{10} & \textbf{11} & \textbf{12} & \textbf{13} & \textbf{14} & \textbf{15} & \textbf{16} & \textbf{17} \\ \midrule
\textbf{x} & 0 & 0 & 0 & 0 & 0 & 0 & 0 & 0 & 0 & 0 & 0 & 0 & 0 & 0 & 0 & 0 & 0 \\
\textbf{y} & 0 & 0 & 2 & 4 & 6 & 8 & 0 & 2 & 4 & 6 & 8 & 0 & 2 & 4 & 6 & 8 & 14 \\
\textbf{z} & 20 & 50 & 50 & 50 & 50 & 50 & 75 & 75 & 75 & 75 & 75 & 100 & 100 & 100 & 100 & 100 & 100 \\
\bottomrule
\end{tabularx}
\end{table*}

\section*{References}
\bibliography{references}

\end{document}